\documentclass[reprint,twocolumn,preprintnumbers,pre,amsmath,amssymb]{revtex4}
\usepackage{dcolumn}
\usepackage{hyperref}
\usepackage{color}
\usepackage{graphicx}
\usepackage{bm}
\usepackage[justification=raggedright]{caption}
\usepackage{amsmath}
\begin{document}
\title{Direct experimental determination of critical disorder in one-dimensional weakly disordered photonic crystals}
\author{M. Balasubrahmaniyam}
\affiliation{Nano-optics and Mesoscopic Optics Laboratory, Tata Institute of Fundamental Research, 1, Homi Bhabha Road, Mumbai, 400 005, India}
\author{Sushil Mujumdar}
\email[]{mujumdar@tifr.res.in}
\homepage[]{http://www.tifr.res.in/\~mujumdar}
\affiliation{Nano-optics and Mesoscopic Optics Laboratory, Tata Institute of Fundamental Research, 1, Homi Bhabha Road, Mumbai, 400 005, India}

\date{\today}
\begin{abstract}
We report experimental measurement of critical disorder in weakly disordered, one-dimensional photonic crystals. We measure the configurationally-averaged transmission at various degrees of weak disorder. We extract the density of states (DoS) after fitting the transmission with theoretical profiles, and identify the Lifshitz tail realized by weak disorder. We observe the vanishing of Van Hove singularities and the flattening of the DoS with increasing disorder in our system. Systematic variation of disorder strength allows us to study the behavior of Lifshitz exponent with the degree of disorder. This provides a direct handle to the critical disorder in the one-dimensional crystal, at which the transport behavior of the system is known to change. The contradictory behavior at very weak disorder in the DoS variation at the bandedge and the midgap are seen to resolve into synchronous behavior beyond the critical disorder. The experimentally measured transmission is shown to carry a clear signature of the critical disorder, which is in very good agreement with the theoretically expected disorder.

\end{abstract}
\maketitle

\section{Introduction}

Wave transport in disordered systems is one of the most intriguing fields in physics, as the disorder results in interesting emergent mesoscopic phenomena such as wave diffusion, weak localization, Anderson localization etc\cite{Rotter17}. While diffusion dictates the overall incoherent transport in disordered systems, Anderson localization, which originates from the coherent interference of randomly scattered waves, leads to the arrest of transport\cite{Anderson58}. Localization leads to a disorder-induced metal-insulator transition, and is shown to be ubiquitous across a wide range of systems, ranging from electronic or phononic transport in condensed matter systems, transport in matter-waves, and light transport in disordered photonic systems \cite{Segev13, billy08, wiersma97, raedt89, schwartz07, Lahini08, Pandey16, mookherjea08}.   Following the seminal works of John\cite{John87}, disorder introduced in periodic optical systems has been a topic of intense research, as it provides a roadmap of direct observation of photon localization. Such systems are commonly known as periodic-on-average random systems, PARS in short. In a three-dimensional PARS structure, the states in the passband get localised only beyond a critical degree of disorder. This criticality essentially separates the diffusive states from the localized states. On the other hand, in low-dimensional systems, the states in the passband can be localized even at weak disorder at sufficiently large sample size.

In a PARS structure, disorder leads to the appearance of localized modes close to the bandedge of propagating photonic modes, leading to the modulation of the bandedges. As a result, the sharp bandedges get smeared into an exponentially decaying tail, known as the Lifshitz tail, in the profile of the averaged density of states\cite{Lifshitz64}. The Lifshitz tail essentially comprises the localized modes at frequencies close to the bandedge. As a result, under conditions of weak disorder, the transport not only depends on disorder strength but also depends strongly on the wavelength. This scenario of weakly disordered low-dimensional systems has witnessed intense activity in recent years. Measurements on lifetime statistics of emitters placed in photonic quasicrystals have directly demonstrated the modulation of DoS at weak disorder\cite{Birowosuto10,Krachmalnicoff10, Sapienza11,Ruijgrok10}, including unintentional fabricational disorder\cite{Wang11}. Disorder-induced localized modes close to bandedges have been demonstrated in diverse photonic systems such as photonic crystal waveguides \cite{Topolancik07, Thomas09, Sapienza10, Garcia10, Spasenovic12} and coupled cavity array \cite{Tiwari13, Pandey16}in the limit of weak disorder. Transport properties such as average transmission and localization length are known to vary differently with disorder in the passband and the bandgap. Recently, two dissimilar evolutions of mean localization length with disorder were demonstrated via photoluminescence measurements and simulations in the weak disorder limit\cite{García17}. This wavelength-dependence is, however, broken at a higher disorder. A critical degree of disorder has been defined, beyond which the transport parameters behave congruently in and outside the passband. The critical disorder that is referred to in this work essentially refers to the varying behavior in the passband and bandgap. A signature of criticality was found to be the crossing of the profiles of mean transmittance and variance of transmittance as a function of disorder.\cite{Kaliteevski06} This critical disorder also manifests in the situation where the boundary between states with different properties of localization disappears. In other words, the variance of the Lyapunov exponent starts following its mean only after the critical disorder, before which the two show contrary behavior\cite{Deych98}. Essentially, the critical disorder separates the single parameter scaling regime from the multiparameter scaling.

It has been theoretically shown that this degree of critical disorder is determined by the gapwidth, the density of states (DoS), and the Lifshitz penetration depth\cite{Kaliteevski06}. All these parameters can be identified once the DoS is known. Some techniques of the estimation of DoS have been reported in photonic systems earlier. For instance, in photonic crystal waveguides, the DoS has been reconstructed using the Fourier-transform of the field profile of the modes\cite{Huisman12}, which revealed an exponentially decaying distribution of modes in the bandgap. In several systems, the local density of states has been measured using the lifetimes of emitters coupled to the localized modes \cite{Sapienza10, Birowosuto10,Krachmalnicoff10, Wang11}.  However, to our knowledge, there is no experimental report on the measurement of the critical degree of disorder. This could perhaps be because of the stringent control on weakness of disorder that is required for such a measurement. Indeed, a finely varying disorder overwritten on a periodic template is necessary for such a measurement. In this work, we have experimentally measured this critical degree of disorder in a one-dimensional system. We use the variation of the mean transmittance as an indicator of the critical disorder. We find that this degree of disorder is in excellent agreement with the theoretically-predicted value based on the Lifshitz penetration depth into the bandgap. To this end, we studied the transport in a one-dimensional periodic-on-average random system with finely tunable disorder. The experimentally measured transmittance allowed us to extract the density of states and thus the Lifshitz exponent. The variation of Lifshitz exponent with disorder strength provides the theoretically-expected critical disorder for our system. We show that the experimental mean transmittance shows a marked change in behavior at the critical disorder.

The system under consideration is a linear array of coupled microresonators. Each microresonator is in the form of a liquid microdroplet, created out of a $7:3$ mixture of ethylene glycol (EG) and methanol. The array is generated using a vibrating orifice aerosol generator\cite{Lin90}. The device is based on a periodic perturbation of an unstable liquid jet. A liquid jet formed when exiting a narrow orifice under pressure, breaks up into monodisperse, equi-spaced periodic microdroplets when perturbed periodically with an appropriate frequency and amplitude. In our experiments, a microcapillary (orifice diameter $10~\mu$m) was fitted with a piezo-activated gate that induced periodic perturbations, realizing microdroplets with diameters in the neighborhood of $14-20\mu$m. The array of highly monodisperse and equispaced microspheres constitutes a one-dimensional photonic crystal superlattice. An inherent randomness exists in the microdroplet sizes due to their fluidic nature, leading to an unavoidable, minimum disorder. Further on-demand disorder could be introduced simply by tuning the perturbation frequency. We have earlier reported Anderson localization lasing in the microresonator array under conditions of amplification\cite{randhir17}. The major advantage of this system is the finesse with which the disorder can be tuned, allowing for very fine control of the disorder strength in the PARS structure. Such fine control was crucial to access the critical disorder in the system. In the current experiments, Rhodamine 6G molecules were added to the solvent at a low concentration ($<10^{-4}M$), for two purposes. Firstly, under optical pumping, the emission from Rhodamine excited the Mie resonances which allowed for accurate size measurement of the microresonators. Secondly, under passive conditions, the molecules also acted as local intensity probes via Rayleigh scattering of the propagating light.

\begin{figure}[h]
\includegraphics[width=8cm]{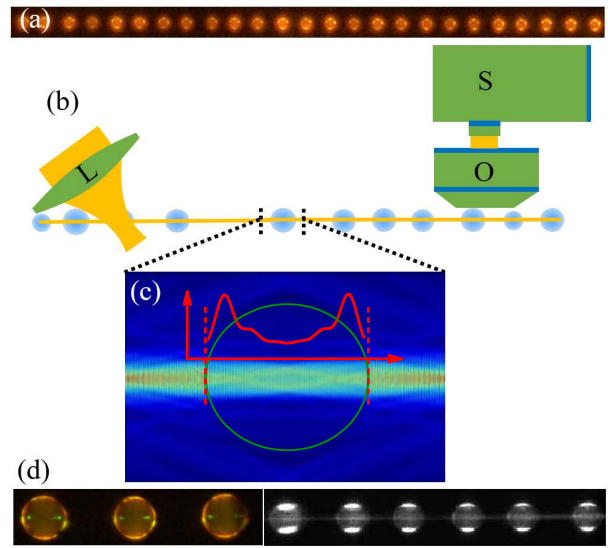}
\caption{\label{fig:wide} (a) Photographic image of the microsphere array. (b) Schematic of the experimental setup. Lens L incouples supercontinuum light into the array.  Scattered light is collected from the array about 60 unit  cells away from the excitation point using the objective O, and spectrally resolved by the Spectrometer S. (c) Intensity distribution inside a single resonator as computed from finite element analysis. Red plot shows envelop of the field at the axis of the array. (d) Magnified true-colour image of three microspheres, showing intensity maxima observed in the finite-element analysis. Monochrome image shows the Mie resonances and the axial mode simultaneously, when excited by pulsed Nd:YAG laser and supercontinuum laser simultaneously.}
\end{figure}

Fig~1(a) shows the photographic image of a part of the array when excited transversely with an Nd:YAG laser (not shown) with pulses of $\lambda = 532.8$~nm. The Mie-resonant emission was spectrally analyzed to obtain an accurate estimate of the individual droplet diameters, and thus the strength of disorder. For the data in this report, the average microsphere diameter was measured to be $\langle d \rangle = 15.5~\mu$m. To measure the transmittance under passive (i.e., zero gain) conditions, we employ a supercontinuum source that is bandpass-filtered to a wavelength range of 670-690~nm. This wavelength range is outside the absorption range of the dye molecules, and hence the molecules only act as passive scatterers. As shown in Fig~1(b), the super-continuum light is injected into the array at an angle of roughly 12$^o$ to the axis, which is within the acceptance angle of the photonic mode of the one-dimensional crystal. The lens ‘L’ provides the required range of wavevector components to facilitate coupling of incident light into the collective mode. An objective ‘O’ was positioned about 60 unit cells away from the injection point. The objective collected the out-of-axis scattered intensity from the modes, which was analyzed with the Spectrometer `S'. The integration time of the detection was maintained 100~ms. Since the supercontinuum source was effectively continuous wave ($\sim40$~MHz repetition rate), the detection process measured configurationally-averaged transmission spectra. Fig~1(c) shows the calculated intensity profile in a single microsphere, as computed from finite-element analysis. (See Supplementary Information.) The red lineplot shows the envelop of the intensity profile along the axis of the array, depicting two intensity maxima inside the microsphere close to the surfaces. The true-color image in Fig~1(d) shows the experimental high-magnification image of three microdroplets, under single-shot illumination from the Nd:YAG laser. Two bright greenish-yellow spots can be seen along the axis of the spheres close to the edges, in agreement with the computation, identifying the photonic mode.  Fig~1(d) also shows a high-magnification, monochrome image of a section of the microresonator array under simultaneous illumination by the Nd:YAG laser and the supercontinuum laser. Here, the equatorial Mie resonances excited by the Nd:YAG pump light are seen together with the collective photonic crystal mode which is excited by the supercontinuum light. By reducing the input slit-width of the Spectrometer, the Mie resonances could be easily masked, thus allowing clean access to the photonic mode. The objective imaged the photonic mode over about 6 microspheres for spectral analysis.

The out-of-axis scatter essentially constitutes the loss in the system, and needs to be quantified for accurate estimation of the DoS. This loss originates from the curved interface between the microsphere and air, and the nonresonant scattering off Rhodamine molecules. The loss is quantified as follows. In the experimental setup, the objective is replaced with the lens which images the collective mode over about 120 resonators onto the CCD, as shown in Fig~2(a). This mode is then spectrally resolved. The inset in Fig~2(b) shows a part of the measured transmission spectrum at the minimum disorder, i.e., the best possible periodicity in the aerosol generator. Two clear bandgaps (centered at $\lambda = 676.43$~nm and $\lambda = 682.71$~nm, are shown. The vertical dotted line in the inset identifies the wavelength $\lambda = 679.5$~nm, which is at the middle of the passband. The main plot in Fig~2(b) shows the decaying intensity profile measured at this wavelength. The profile show a clear exponential decay and the decay length is measured to be $\sim48$ unit cells. In general, the decay in a disordered system is manifested by two processes, namely, (a) the disorder-induced Anderson localization, and (b) the systemic losses. Given the localization length $\xi$, the effective decay length is given $[1/\xi+1/\ell_{loss}]^{-1}$. However, it is known that $\xi\rightarrow\infty$ at frequencies in the middle of the passband at low disorder strengths. Thus, the $\xi >> \ell_{loss}$. As a result, the effective decay is dominated by the $\ell_{loss}$. We note that the droplet curvature and refractive index is uniform over the narrow range of wavelengths studied in this work. Hence, the $\ell_{loss}$ is assumed constant with respect to wavelength.

\begin{figure}[h]
\includegraphics[width=8cm]{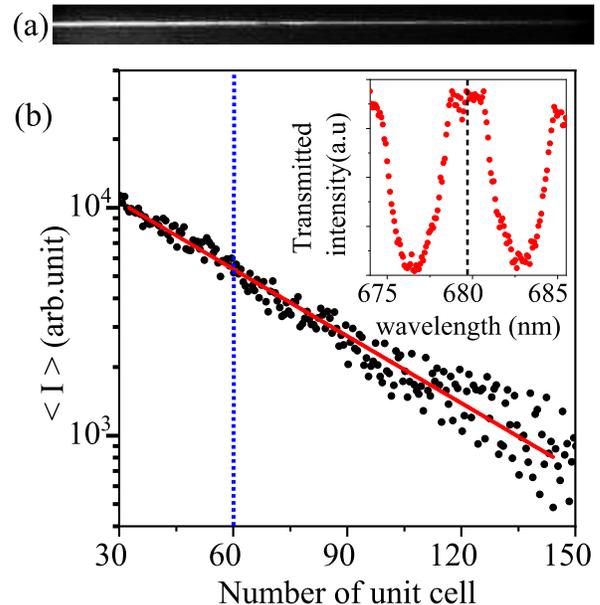}
\caption{\label{fig:epsart} (a) Image of the decaying photonic mode as taken by a CCD. (b) Inset shows the measured transmission spectrum at the weakest disorder, showing two bandgaps. Vertical line marks the passband center. Main plot shows the intensity decay at the passband center, with the red solid line the exponential fit giving the decay length to be $\sim48$ unit cells.}
\end{figure}

The averaged transmitted intensity of the collective mode is measured for different disorder strengths, which is realized by detuning the piezo-gate frequency of the vibrating orifice aerosol generator. The spectra are measured at a distance of about 60 unit cells from the point of illumination. The objective images a group of about 6 microresonators onto the spectrometer, which resolves the collective mode into its spectral components. The spectra of the transmitted intensity measured at different disorder strengths ($\delta$) are provided in Fig. 3(a). The disorder $\delta$ is quantified as $\delta=n_d\sigma_d + \sigma_a$, where $n_d$ and $\sigma_d$ are the refractive index and the standard deviation in the size of the dielectric (in this case, the droplet), and $\sigma_a$ is the standard deviation in the air gap between the droplets. Essentially, $\delta$ is the deviation in a unit cell, measured in nm. The low and high transmitted intensity regions in the spectra show the transmission bands and bandgaps, respectively, of the 1D photonic crystal realized by the array. The band edges show a clear change in slope with varying disorder strength and the transmitted intensity profile gets flatter at higher disorder strengths.

\begin{figure}[h]
\includegraphics[width=8cm]{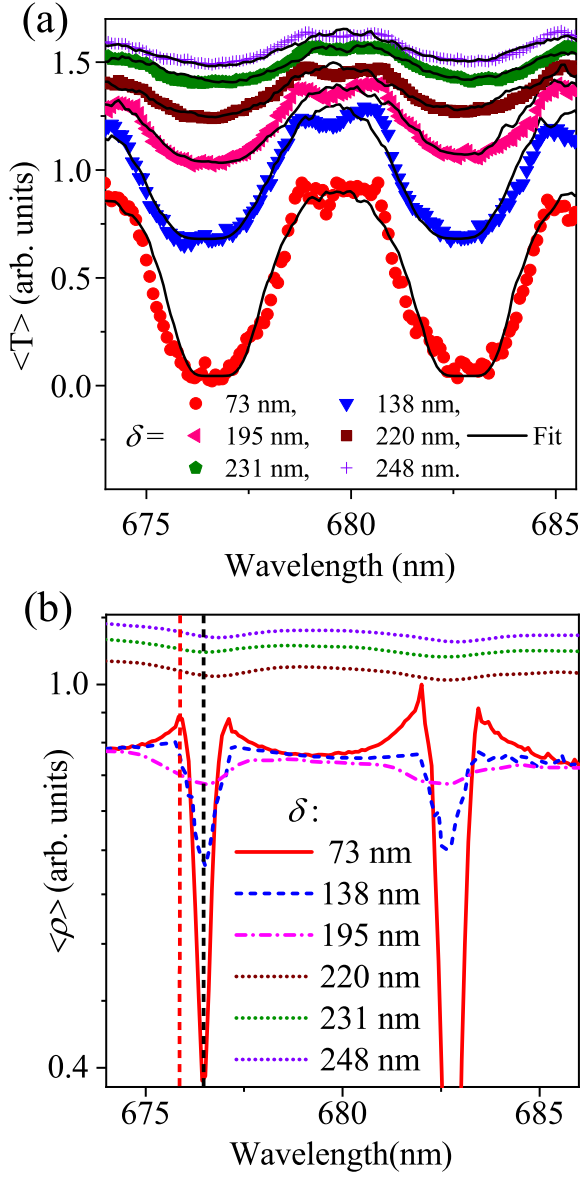}
\caption{\label{fig:epsart} (a) Coloured markers depict the transmitted intensity spectra for various disorder strengths, measured using the setup in Fig~1(b). Black curves show least-squares fits. Legend identifies the disorder strength $\delta$.(b) Density of states ($\langle \rho(\lambda) \rangle$) extracted from the experimentally measured $\langle T \rangle)$. Three dotted lines (at $\delta$ = 220, 231, \& 248~nm) are vertically offset for visibility. Vertical red dashed line marks the bandedge, while the black dashed line shows the gap center.}
\end{figure}

Next, in order to extract the density of states, we first estimate the complex transmittance using the 1-D transfer matrix method (TMM) embedded into a Levenberg-Marquardt algorithm\cite{levenberg44,marquardt63}. The TMM simulates the experimental system as a binary multilayer, with alternating refractive indices of $n_d = 1.4$ (for EG-Methanol mixture) and 1 (for air). To simulate losses, imaginary component $\kappa_d$ and $\kappa_a$ were added to the refractive indices of both, the dielectric and air layers. The complex refractive index for the air layer is plausible because the curved interfaces in the experimental system scatter both the dielectric modes and the air modes. Hence, the experimental transmission spectra exhibit comparable amplitudes for the dielectric and air bands. If the loss is only included in the dielectric, it leads to a diminished transmission of the dielectric band in comparison to the air band. The number of unit cells was taken to be 60, comparable to the experimental sample length over which the transmission was measured. For the near-periodic (minimum disorder) system, the multiparameter fit implemented by the Levenberg-Marquardt algorithm began with an initial guess  $\{\langle d \rangle^{0}, \langle a \rangle^{0}, \sigma_d^{0},\sigma_a^{0},\kappa_d^{0},\kappa_a^{0}\}$, where $\langle a \rangle$ was the mean inter-resonator spacing. $\langle a \rangle^{0}$ and $\sigma_a^{0}$ were taken equal to $\langle d \rangle^{0}$ and $\sigma_d^{0}$, as measured from photographic images of the resonator array (Fig~1(d)). The other parameters were set as determined from the experiments. $\kappa_d^0 = \kappa_a^0 = 1/2\ell_{loss}$ was set as determined by the loss length as computed in Fig~2. Note that the number of layers was not a fit parameter, as the decay coefficient of the Lifshitz tail does not depend on the sample size. (See Supplementary Information). First, the algorithm computed the average transmission spectrum $\langle T \rangle^0$ over a 1000 configurations. Next, the algorithm iteratively varied the parameter-set to provide the bestfit values of \{$\langle d \rangle^n, \langle a \rangle^{n}, \sigma_d^{n},\sigma_a^{n},\kappa_d^{n},\kappa_a^{n}$\} after $n$ iterations, thus yielding $\langle T \rangle^n$, the bestfit to the experimentally measured spectrum. We found that the $\ell_{loss}$ turned out to be independent of disorder strength in this analysis. This is plausible since the number of scatterers does not change over the disorder strengths within the small range of disorder considered herein. Note that, the bestfit $\langle d \rangle^{n}$ and $\langle a \rangle^{n}$ obtained for the near-periodic structures were used as fixed parameters for higher disordered samples. The black curves in the Fig~3(a) show the fits. The legend in the figure shows the $\delta$ representing the bestfit disorder parameters.

Once the bestfit parameters were obtained, the complex transmission spectrum was computed using \{$\langle d \rangle^n, \langle a \rangle^{n},\sigma_d^{n},\sigma_a^{n},\kappa_d^{n},\kappa_a^{n}$\}, for another set of 1000 spectra. From the phase part of each transmittance used in the fit, the DoS $\rho(\lambda)$ for individual configurations were reconstructed\cite{Avishai85}. Fig.~3(b) shows the $\langle\rho(\lambda)\rangle$, averaged over 1000 configurations. At the lowest disorder strength ($\delta = 73$~nm, solid red curve), the $\langle\rho(\lambda)\rangle$ show a remnant of the Van Hove singularity that exists at the band edges in the periodic limit. At larger disorder strengths, the peak at the bandedge disappears ($\delta = 138$~nm, dashed blue curve). For the disorder of $\delta = 195$~nm (dash-dotted magenta curve), the dip in $\langle\rho(\lambda)\rangle$ at the bandgap is significantly reduced leading to an almost flattened profile. The curves for higher disorder, for $\delta = 220, 231$ and 248~nm, are offset vertically for visibility, and are seen to be flat. The vertical red dotted line marks $\lambda = 675.84$~nm, close to a band edge. The black dotted line marks $\lambda = 676.43$~nm, which is in the middle of the gap.

\begin{figure}
\includegraphics[width=8cm]{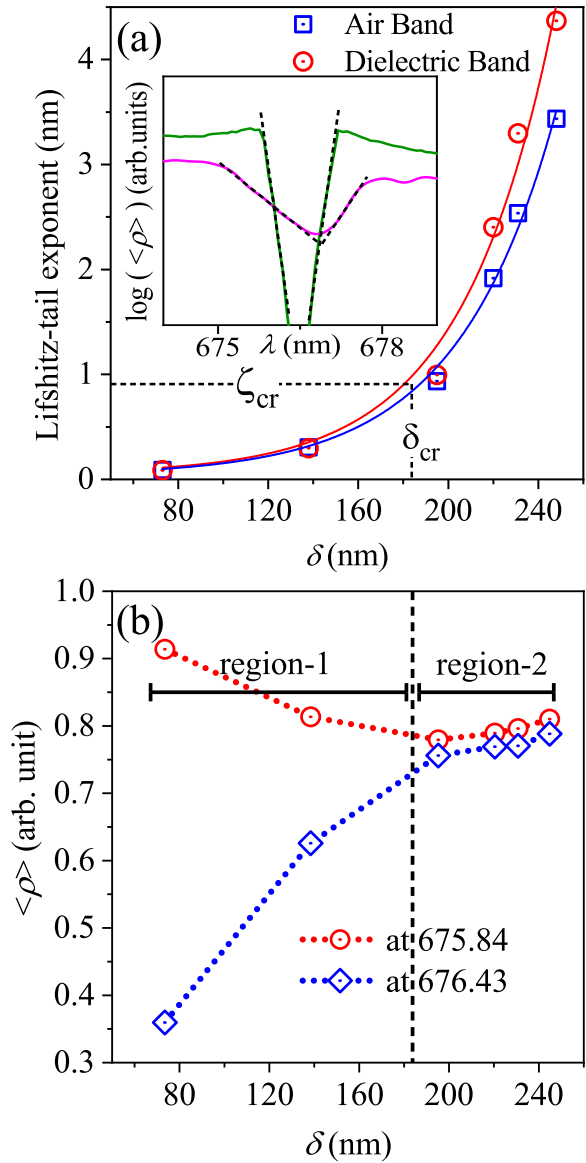}
\caption{\label{fig:epsart} (a) Lifshitz tail exponent $\zeta$ as function of disorder strength $\delta$, for the air band (red circles) and the dielectric band (blue squares). The line is a guide to the eye. Inset: $\rho(\lambda)$ for two disorder strengths $\delta = 138$~nm (green line) and $\delta = 221$~nm (magenta line) on log Y axis, with the exponential fits. Horizontal dotted line marks the critical tail exponent $\zeta_{cr} = 0.91$~nm, which corresponds to $\delta_{cr} = 183$~nm (vertical dotted line). (b) Variation of $\langle\rho(\lambda)\rangle$ with $\delta$, for bandedge wavelength of 675.84~nm (red circles) and midgap wavelength of 676.43~nm (blue diamonds). Upto $\delta= \delta_{cr}$,  an opposite behavior is seen, while beyond $\delta_{cr}$, they vary synchronously.}
\end{figure}

The tail in the $\langle\rho(\lambda)\rangle$ at the bandedges exhibits an exponential decay $\exp [-(\lambda - \lambda_0)/\zeta]$, where $\zeta$ is the Lifshitz tail exponent, and $\lambda_0$ is the bandedge wavelength. Fig~4(a) depicts the behavior of $\zeta$ with $\delta$.  For the bandgap centered at $\lambda = 676.5~$nm, the inset shows the $\langle\rho(\lambda)\rangle$ at two disorder strengths, namely, $\delta = 138$~nm (green curve) and $220$~nm (magenta curve). The semi-logarithmic axes emphasize the exponential decay, whose fit provides the $\zeta$. The left edge belongs to the dielectric band, while the right edge arises from the air band, as identified in the transfer matrix computations. As expected from earlier data, an asymmetry is seen to exist in the two edges which is endorsed and quantified in the main plot. The $\zeta$ varies from $\sim0.1$~nm at $\delta = 73$~nm to $\sim4$~nm at $\delta = 240$~nm. The $\zeta$ for the air band (blue $\square$'s) is consistently smaller than that for the dielectric band (red $\circ$'s).

The critical disorder $\delta_{cr}$ is revealed in the change of behavior of average transmission $<T>$ or the Lyapunov exponent $\bar{\lambda}$. (See Supplementary Information.) The $\delta_{cr}$ has been shown to depend on the gapwidth $\Delta\lambda_g$ and the central wavelength of the gap $\lambda_g$. If $S_{th}$, the Suppression factor, is the ratio of the DoS at the gap center to that at the gap edge, then the Lifshitz tail coefficient $\zeta$ at $\delta_{cr}$ works out to be\cite{Kaliteevski06}
\begin{equation}
\frac{\zeta_{cr}}{\lambda_g} = \frac{1}{2\sqrt{-\ln(S_{th})}}\frac{\Delta\lambda_g}{\lambda_g}
\end{equation}
In our experiments, the $\Delta\lambda_g = 2.5$~nm, the $S_{th} = 0.15$, and the $\zeta_{cr}$ was estimated to be 0.91~nm. This value is marked by the horizontal dashed line in Fig~4(a). The corresponding $\delta = \delta_{cr} = 183$~nm, identified by the vertical dashed line.

The variation of $\langle\rho\rangle$ with the disorder at the bandedge and midgap wavelengths is shown in Fig~4(b). The red $\circ$'s indicate that the $\langle\rho\rangle$  at the band edges initially drop with disorder, and then increase again after $\delta_{cr}$. The blue $\lozenge$'s show that the DoS in the middle of the gap consistently increases with $\delta$, and after  $\delta_{cr}$, the two curves grow in parallel. The two dissimilar functional dependencies are demarcated by the black vertical dashed  line at $\delta_{cr}$. Notably, in region 2, there is a small constant difference in the $\langle\rho\rangle$  of $\sim0.02$ (in normalized units), which originates from the fact that the response of the dielectric-bands and air-bands to disorder are slightly different\cite{Cao10}. This was already quantified in the Lifshitz tail, as shown in Fig~4(a).

\begin{figure}
\includegraphics[width=8cm]{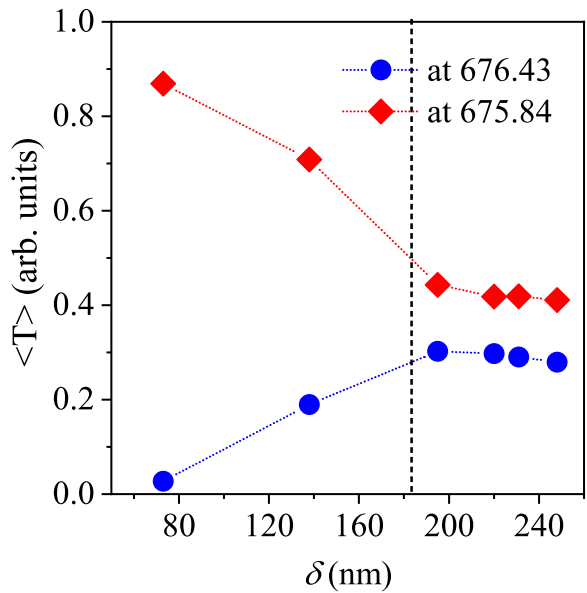}
\caption{\label{fig:epsart} Experimentally measured $\langle T \rangle$ with $\delta$ at the two wavelengths, showing a clear change in behavior at $\delta_{cr}$. The dashed line shows the calculated $\delta_{cr}$ using the Lifshitz-tail exponents. }
\end{figure}

Finally, we verify the consequence of $\delta_{cr}$ on the transmission. Fig~5 depicts the experimentally measured $\langle T \rangle$ as a function of $\delta$ at the two wavelengths under discussion. The red $\blacklozenge$ shows a rapidly decreasing behavior upto the $\delta = 195$~nm, after which the decrease is gradual. On the other hand, at the midgap ($\lambda = 676.43$~nm, blue $\bullet$'s), the $\langle T \rangle$ increases rapidly for weaker disorder, and at 195~nm, this rise also slows and the $\langle T \rangle$ for the two wavelengths varies in parallel. Thus, a clear change in $\langle T \rangle$ is observed experimentally in the weakly disordered one-dimensional photonic crystals. The vertical dashed line shows the theoretically obtained $\delta_{cr}$ at 183~nm, which is in excellent agreement with the experimentally measured data at 195~nm. While the Lifshitz tail coefficient identifies the $\delta_{cr}$, this is also a confirmatory measurement of the same.
We emphasize that for layer thicknesses of about $\sim15~\mu$m, the achieved variations of a few tens of nm's is a very fine variation, allowing us access to $\delta_{cr}$.

We further note that our experimental system has a finite loss, which, as reported in the results, does not affect the critical degree of disorder. This allows us a handle on investigating the phenomenon of Anderson localization under weak dissipation, where the localization properties below and above the critical degree of disorder can be studied. Although dissipative Anderson localizing systems are antipodal to amplifying Anderson localizing systems, they have not seen comparable research clearly due to the detrimental nature of dissipation. Weakly-dissipative or controlled-dissipative structures can provide significant insights into physics of Anderson localization\cite{balu2018}.

In conclusion, we report the direct experimental measurement of critical disorder in a one-dimensional photonic crystal. The experimental setup with a fine control on disorder strength allows us to measure variations in transport over incremental increases in disorder strength. We reconstruct the ensemble-averaged density of states of the photonic modes from the transmission measurements. The remnant of Van Hove singularity is observed at the bandedge at very weak disorder. With increasing disorder, we observe the Lifshitz tail in the DoS profile, which eventually flattens out and loses wavelength dependence. The systematic evolution of the Lifshitz exponent with disorder enables us to identify the critical degree of disorder. The evolution of DoS with disorder strength at the bandedge and the gap clearly show dissimilar functional dependencies below the critical disorder strength, beyond which they monotonically and parallely increase with disorder strength. The experimentally measured average transmission shows a clear change in behavior at the theoretically estimated critical degree of disorder.

\begin{acknowledgments}
We wish to acknowledge funding from the Swarnajayanti Fellowship Grant, Department of Science and Technology, Government of India.
\end{acknowledgments}

\end{document}